\documentclass[fontsize=11pt,a4paper,toc=flat]{arxiv-article}
\usepackage{siunitx}
\usepackage{Definitions}

\DeclarePairedDelimiterXPP\mudim[1]{}\llbracket\rrbracket{_\mu}{#1}

\newcommand*{\Op}{\mathcal{O}}
\usepackage{tikz}
\usetikzlibrary{arrows.meta, decorations.pathreplacing}
\usepackage{hyperref}
\graphicspath{{./plots/}{./FIGS/}}

\preprint{\texttt{NT@UW-26-15}}

\acrodef{lec}[\textsc{lec}]{low-energy constant}

\newcommand{\OfficialTitle}{
  The odd fermion at the edge:\\
  odd-even staggering in the trapped, unitary Fermi gas
}
\title{\setstretch{1.4}
	{\color{Thoughtless}\textls[-20]{\OfficialTitle}}
}

\hypersetup{pdfauthor={Silas R.~Beane, Domenico Orlando, Susanne Reffert},pdftitle={\OfficialTitle},%
	colorlinks=true,linkcolor=ThoughtYouWere,citecolor=ThoughtYouWere,urlcolor=ThoughtYouWere}

\author{%
	\begin{minipage}{.94\textwidth}
		\begin{center} \dosserif%
			{\small
           \textbf{Silas~R.~Beane}\textsuperscript{\ding{74}},
				\textbf{Domenico~Orlando}\textsuperscript{\ding{72}\ding{73}}, and
  				\textbf{Susanne~Reffert}\textsuperscript{\ding{73}} 
			}
		\end{center}
  		\authorBlock{\ding{74}}{\dosserif{}%
                        Department of Physics,\\
                        University of Washington,\\
                        Seattle, WA 98195}
		\authorBlock{\ding{72}}{\dosserif{}%
			INFN sezione di Torino.\\
			via Pietro Giuria 1, 10125 Torino, Italy}
		\authorBlock{\ding{73}}{\dosserif{}%
			Albert Einstein Center for Fundamental Physics,\\
			Institute for Theoretical Physics, University of Bern,\\
			Sidlerstrasse 5, CH-3012 Bern, Switzerland}
	\end{minipage}
}

\date{}

\begin{document}

\numberwithin{equation}{section}

\begin{titlepage}

  \maketitle

  \thispagestyle{empty}

  \vfill

  \abstract{\normalfont{}\noindent{} 
We investigate the odd-even staggering in the harmonically-trapped unitary Fermi gas at large particle-number charge $Q$.
Using both a large-$N$ \acs{bdg} description and a complementary large-charge \acs{eft} method, we show that for odd particle number the extra fermion forms an edge-localized quasiparticle near the Thomas--Fermi surface rather than a bulk excitation.
In the edge limit, the microscopic \acs{bdg} problem reduces to a universal coupled Airy system whose lowest positive eigenvalue fixes the leading odd-even splitting energy,  $\chi\,\xi^{1/6}(24Q)^{1/9}\,\hbar\omega + \cdots$ where $\xi$
    is the Bertsch parameter, and $\chi$ is a universal edge
    coefficient.
The associated \acs{eft} describes a fermionic mode confined to the boundary and coupled to the superfluid Goldstone field, reproducing the same $Q$ scaling while introducing a dependence on two low-energy constants.
Finally, we numerically compute the spectrum and confirm the predicted scaling and localization properties.
  }
  \vfill

\end{titlepage}

\setstretch{1.1}
\tableofcontents

\section{Introduction}

The unitary Fermi gas is a universal many-body system in which
two-component fermions interact with infinite $s$-wave scattering
length and negligible effective range and higher-order shape
parameters.  In this regime the microscopic details of the interaction
are washed out, and the many-body problem is controlled by symmetry,
dimensional analysis, and a small set of universal numbers.  This
makes the unitary gas a useful bridge between nuclear many-body
physics, and strongly-interacting cold atomic gases~\cite{Giorgini_2008}.  Experimentally, the system is
realized with ultracold atoms tuned close to broad Feshbach
resonances, while theoretically it provides a stringent benchmark for
\ac{qmc} methods, density functional theory, \ac{eft}, and \ac{bdg} methods.

A central universal quantity which describes the unitary Fermi gas is
the Bertsch parameter $\xi$, defined by the ground-state energy
density of the homogeneous unpolarized unitary gas relative to that of
the free Fermi gas at the same density.  Equivalently, in a harmonic
trap, $\xi$ controls the leading (Thomas--Fermi) contribution to the
large-$Q$ ground-state
energy~\cite{Son:2005rv,Ma_es_2009,Kravec:2018qnu,Favrod:2018xov}.  Determining
$\xi$ accurately has been a long-standing challenge for both
experiment and numerical simulation, largely because it requires
controlled treatment of a strongly-coupled fermionic many-body system
with no intrinsic length scale other than the interparticle
spacing. At unitarity the system has a remarkably-rich symmetry
structure: it realizes a \ac{nrcft} with
Schrödinger symmetry~\cite{Hagen:1972pd,Niederer:1972zz,Henkel:1993sg,Henkel:2003pu}.  In a harmonic trap this symmetry implies a
non-relativistic version of the state--operator correspondence,
relating scaling dimensions of primary operators in flat space to
energy eigenvalues in the trap~\cite{Nishida:2007pj}.  The trapped gas
therefore provides both an experimentally accessible many-body system,
and a finite-density laboratory for non-relativistic conformal
dynamics.

\begin{figure}
  \centering
  \begin{tikzpicture}[font=\footnotesize]
    \node [above right,inner sep=0] (image) at (0,0) {\includegraphics[width=.5\textwidth]{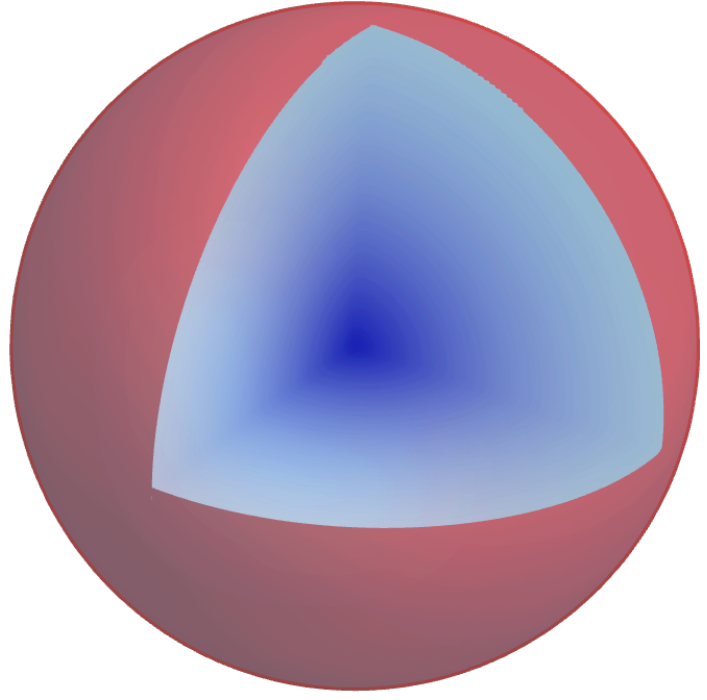}};
    \draw[{Circle[fill=black]}-]  (1,3) -- (-1,1.5) node[below, text width=3cm, align=center] {unpaired \(l=0\) fermion at the edge};
    \draw[{Circle[fill=black]}-]  (4,3) -- (8,1) node[below, text width=3cm, align=center] {paired fermions (superfluid bulk)};
  \end{tikzpicture}
  \caption{The unpaired fermion localizes at the edge of the cloud, where the gap is minimal. Its ground state is a homogeneous \(l = 0\) configuration.}
  \label{fig:cloud-in-the-shell}
\end{figure}

Odd fermion particle number adds a further universal question.  For
even $Q$, the ground state is naturally interpreted as a paired
superfluid configuration.  For odd $Q$, all but one of the fermions
can pair, leaving a single unpaired quasiparticle in the trap.  This
produces an odd--even staggering of the ground-state energy, analogous
in spirit to odd--even effects in finite nuclei.  The trapped unitary
gas is therefore a clean universal system for asking how a single
fermionic quasiparticle behaves in the background of a strongly paired
many-body system.  In Ref.~\cite{Son2007}, Son argued that, at large
odd $Q$, the relevant quasiparticle is not localized at the center of
the cloud or droplet, where the pairing gap is largest, but rather
near the edge, where the local gap becomes small and the \ac{lda} breaks down~\cite{Son2007} (see Figure~\ref{fig:cloud-in-the-shell}).
The resulting edge
problem implies a parametrically-small excitation energy compared with
the central pairing gap.
The odd--even splitting, defined, for \(Q\) odd, as
\begin{equation}
  \Delta_{\text{stag}}(Q) = E(Q) - \frac{1}{2} \pqty*{E(Q-1) + E(Q+1) }  ,
\end{equation}
where \(E(Q)\) is the energy of \(Q\) fermions at unitarity in a harmonic trap, is expected to scale as
\begin{equation}
  \label{eq:Son-formula}
  \Delta_{\text{stag}}(Q)
  = \chi\,\xi^{1/6}(24Q)^{1/9}\,\hbar\omega + \cdots,
  \qquad Q\gg 1,\quad Q\ \text{odd},
\end{equation}
where $\xi$ is the Bertsch parameter and $\chi$ is a universal dimensionless coefficient associated with the lowest edge quasiparticle mode.

The same quantity is used in a different physical configuration to compute the pairing-gap observable discussed in radio-frequency spectroscopy experiments~\cite{Z_rn_2013,Chin_2004}.
In presence of the trap, the fermion lives at the edge and its ground state energy probes the boundary physics.
For a homogeneous system, instead, the odd fermion diffuses in the full system and its energy probes the bulk physics.
In this work we will concentrate on the confined case.

\medskip
The purpose of this work is to give a more-systematic basis to Son's
edge-quasiparticle argument by making use of droplet-edge \ac{eft}
methods~\cite{Hellerman:2020eff}, and to provide an estimate of the
universal Son parameter $\chi$ using a \ac{bdg} treatment %
that can in turn be
compared directly with existing few-body and quantum Monte Carlo
data~\cite{Chang_2007,Blume_2007}.

We use a three-pronged line-of-attack:
\begin{itemize}
\item First we use a probe approximation, in which the
  odd fermion is a quasiparticle localized at
  the edge of a particle cloud, and
  is the lowest mode for a \ac{bdg} Hamiltonian
  in the \acf{lda}, expanded around the edge.
  Probe means that there is no backreaction of the odd fermion on the underlying condensate.
  To relax this assumption,
\item we write an
  \ac{eft} for the condensate augmented by a coupled fermion edge term.
  The only assumption is that the fermion lives at the cloud edge.
  In this way we confirm the previous estimates for the parametric dependence of the quasiparticle energy, and we can derive the explicit coupling between the fermion and the bulk Goldstone mode.
  However, the result depends on a set of low-energy constants that have to be computed independently.
  To do this,
\item we perform a numerical analysis, again in the probe approximation, but without assumptions regarding
  the localization of the fermion.
  We use the full \ac{bdg} Hamiltonian and compute its low-lying spectrum.
\end{itemize}
For the \ac{eft} treatment, we apply the large-charge approximation to the unitary Fermi gas. In a sector of fixed global charge, which is taken to be large, an \ac{eft} is constructed as an expansion in negative powers of the large charge $Q$~\cite{Hellerman:2015nra,Gaume:2020bmp}. While this method was first developed for the (relativistic) $O(N)$ vector model, it lends itself also to other systems which can be described by a non-relativistic superfluid, such as the Fermi gas which at its unitary point is described by a \ac{nrcft}~\cite{Favrod:2018xov,Kravec:2018qnu,Orlando:2020idm,Hellerman:2021qzz,Hellerman:2023myh,Beane:2024kld,Beane:2025hon,Beane:2025tum}. In this case, the charge $Q$ corresponds directly to the particle number.

The upshot of our analysis is that when a fermion is added to a system of an even number of confined particles, it localizes around the edge of the droplet within a region of order \(\ell_{\delta}\).
  Its low-energy spectrum, up to higher-order corrections, has the form
  \begin{equation}
    \mathcal{E}_{n, l} = \frac{1}{\ell_{\delta}^2} \chi_n + \frac{1}{2 M_{\parallel} R_{\text{cl}}^2} l ( l + 1) , 
  \end{equation}
  where \(\chi_n\) is a set of dimensionless numbers, and \(\ell_{\delta}\) and \(M_{\parallel}\) depend parametrically on the chemical potential and the harmonic trap as follows:
  \begin{align}
    \frac{1}{\ell_{\delta}^2} &= \order*{ \omega Q^{1/9}} ,\\
    M_{\parallel}  &= \order*{ Q^0 } .
  \end{align}
In the microscopic edge analysis below, the universal parameter $\chi$ will be identified with the lowest positive eigenvalue of the universal transverse edge problem.
In the \ac{eft} description, the same leading coefficient is encoded in the \ac{lec} $e_0$.

\bigskip
The plan of this work is as follows. Section~\ref{sec:probe-approximation} studies the extra fermion first in a probe approximation, where the background superfluid is fixed and the unpaired fermion does not backreact on it.
In that limit, the Hamiltonian separates into a motion perpendicular to the edge and a motion along the spherical boundary, producing a low-energy spectrum with edge-localized modes and an effective surface mass.

Section~\ref{sec:eft} rebuilds the same physics using the large-charge \ac{eft}, where the low-energy degrees of freedom are the superfluid Goldstone mode plus a fermion confined to the edge, and it recovers the same parametric scaling for the fermion ground state energy while introducing the \ac{lec}s $e_0$ and $M_{\parallel}$.

Section~\ref{sec:numerics} explains how to solve the trapped radial \acs{bdg} equations numerically using a finite-difference discretization and an Arnoldi eigensolver aimed at the low-lying near-zero modes.
We check whether the states are truly edge-localized by checking that the wavefunction is centered near the classical radius and that its width follows the Airy-layer scaling $\ell_{\delta} \sim (\omega^2 \mu)^{-1/6}$, while the lowest energies scale like $E \sim \omega(\mu/\omega)^{1/3}$.
Our fit table shows exponents very close to these expectations, which supports the edge-quasiparticle picture and Son’s scaling argument.
\section{Probe approximation}
\label{sec:probe-approximation}

In this section, we spell out and make precise the reasoning in~\cite{Son2007}.
We start by considering the problem of the unpaired single fermion in the probe approximation.
A single fermion is added to a system of an even number of fermions that have formed Cooper pairs and have condensed into a superfluid confined in a harmonic potential.
In the probe approximation we neglect the backreaction of the extra fermion on the superfluid.
This approximation is well justified if there are many fermions in the bulk (large charge limit).
We further make use of the \ac{lda} and assume that the unpaired fermion is localized at the edge.

We start from a paired superfluid in a harmonic trap, to which a single fermion is added.
The odd fermion can be described as a Bogoliubov quasiparticle, \emph{i.e.} an eigenstate of the 
\ac{bdg} Hamiltonian~\cite{fetter1971quantum}:
\begin{equation}
  \label{eq:BdG-Hamiltonian}
  H =
  \begin{pmatrix}
    -\frac{1}{2}\Laplacian{} - \mu + V(x) & \sigma(x) \\
    \sigma(x) & \frac{1}{2}\Laplacian{} + \mu - V(x)
  \end{pmatrix},
\end{equation}
where \(V = \frac{\omega^2}{2} r^2\) is the confining potential, %
and the gap \(\sigma(r)\) is the solution to a bulk variational problem.
For \( V = 0\), the gap is constant and proportional to the chemical potential, \(\sigma = y \mu\).
Here, \(y\) is a parameter that is independent of the details of the microscopic physics and can be computed, \emph{e.g.} in the large-\(N\) limit for a system with \(N\) fermionic flavors~\cite{Veillette:2007zz},
\begin{equation}
  y_N = 1.162 + \frac{0.3379}{N} + \order*{\frac{1}{N^2}} \, .
\end{equation}
or via \ac{qmc} simulations~\cite{Carlson:2007omx,Carlson:2014pxa}:
\begin{equation}
  y_{\ac{qmc}}  = 1.22(5) \, . 
\end{equation}
(See also Table~\ref{tab:gap-mu-y}.)
The effect of the potential \(V(x)\) can be taken into account systematically in the large-charge expansion, where the small parameter is \(\omega / \mu\), which is also the small parameter that controls the superfluid \ac{eft} description of Refs.~\cite{Son:2005rv,Ma_es_2009}.
It amounts to a gradient expansion~\cite{Hellerman:2023myh}, leading to
\begin{equation}
  \label{eq:sigma-gradient-expansion}
  \sigma(x) = \begin{cases} y \pqty*{\mu - V(x)} + y_1 \frac{(\nabla V)^2}{\pqty*{\mu - V(x)}^2} + y_2 \frac{\Laplacian V}{\mu - V(x) } + \dots & \sigma > 0\\
   0 & \text{otherwise.}
  				\end{cases} 
\end{equation}
The leading term in this expansion is the so-called \ac{lda}.
The coefficients \(y_1\) and \(y_2\) have been computed at large \(N\) in~\cite{Hellerman:2023myh}:
\begin{align}
  y_1 &= -0.004347 + \order*{\frac{1}{N}}, & y_2 &= -0.1608 + \order*{\frac{1}{N}} .
\end{align}
In the following, we will neglect these higher-order corrections and keep
\begin{equation}
  \label{eq:sharp-edge}
  \sigma(r) = y \pqty*{ \mu - V(r)} \Theta(R_{\text{cl}} - r),
\end{equation}
where \(R_{\text{cl}}\) is the classical radius of the cloud, \emph{i.e.} the region where the particles are confined, defined by \(\mu = V(R_{\text{cl}})\).
This is the scale that controls bulk effects in the theory.
For the harmonic trap, \(R_{\text{cl}} = \sqrt{2 \mu}/ \omega\).

The gap becomes minimal at the classical radius $r=R_{\text{cl}}$, and one expects the unpaired fermion to localize around the cloud edge.
In this region, the gradient expansion breaks down and one needs to zoom in on the edge region (the \(\delta\) layer of~\cite{Orlando:2020idm}) to obtain a different expansion, controlled by a different scale.
We introduce a dimensionless  coordinate \(u\) and a small parameter \(\delta\),
\begin{equation}
  R_{\text{cl}} - r  =  \pqty*{R_{\text{cl}} \delta} u,
\end{equation}
so that the interior of the droplet is in the region $u> 0$.
The precise value of $\delta$ will be fixed by requiring the kinetic and potential terms in \(H_0\) to be of the same order.
This is the ``distinguished limit'' of boundary-layer theory.

The Laplacian on \(\setR^3\), up to terms of order $\delta$, separates into the Laplacian on \(\setR_u\) and the one on the two-sphere of radius \(R_{\text{cl}}\):
\begin{equation}
  \Laplacian_{\setR^3}
  =
  \frac{1}{\delta^2 R_{\text{cl}}^2}\frac{\dd^2{}}{\dd u^2}
  +
  \frac{1}{R_{\text{cl}}^2}\Laplacian_{S^2}{}
  +
  \order*{\frac{1}{R_{\text{cl}}^2\delta}},
\end{equation}
where \(\Laplacian_{S^2}\) is the Laplacian on the unit sphere.
From this formula we also read that at this order the geometry close to the edge is $\setR \times S^2$. 
Similarly, the potential term becomes
\begin{equation}
  \mu - V(r) = \pqty{2 \mu \delta } u + \order{\delta^2} \, .
\end{equation}
We are interested in discussing the large-charge regime \(\omega / \mu \ll 1\).
In order to retain the relevant physical effects, we need to take a distinguished limit \(\delta \to 0\) such that the kinetic and the potential term have the same parametric scaling around the edge.
This happens when
\begin{equation}
  \delta =  \pqty*{ \frac{\omega}{2\mu}}^{2/3} \, .
\end{equation}
It is very important that the model is at unitarity and has no intrinsic scales.
This is also the reason why the new expansion parameter can only depend on the ratio \(\omega / \mu\).

Around the edge, the \ac{bdg} Hamiltonian becomes
\begin{equation}
  H = \pqty{ 2\mu \omega^2}^{1/3} \bqty*{ - \pqty*{ \frac{1}{2}\frac{\dd^2{}}{\dd u^2} + u} \boldsymbol{\tau}_3 + u \ \Theta(u) \sigma_1 } + \frac{1}{2 R_{\text{cl}}^2} \Laplacian_{S^2}{} \boldsymbol{\tau}_3 = \frac{1}{\ell_{\delta}^2} H_{\perp} + \frac{1}{2 R_{\text{cl}}^2} H_{\parallel},
\end{equation}
where \(\boldsymbol{\tau}_3\) and \(\boldsymbol{\tau}_1\) are Pauli matrices.
The distinguished limit identifies a new length scale \(\ell_{\delta}\), which is the typical scale at which edge effects become important:
\begin{equation}
  \ell_{\delta} = R_{\text{cl}} \delta = \pqty*{ \omega \sqrt{2 \mu}}^{-1/3} \ .
\end{equation}
Via the relation
\begin{equation}
	\frac{\mu}{\omega}=\xi^{1/2} (3Q)^{1/3},
\end{equation}
we see that we reproduce the scaling argued in~\cite{Son2007}.

In this limit, the Hamiltonian decomposes into an Airy-type problem for the direction orthogonal to the edge and a motion on the sphere along the edge.
The two components are identified by the natural scales \(\ell_{\delta}\) and \(R_{\text{cl}}\) which ultimately depend only on the external potential via \(\omega\), and the chemical potential \(\mu\) as fermions at unitarity do not have intrinsic scales.

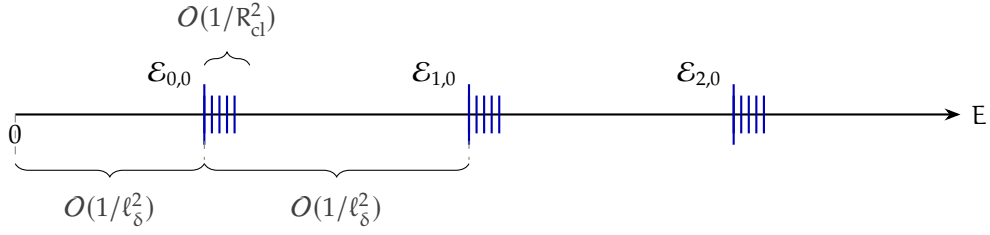
\begin{figure}
  \centering
  \begin{tikzpicture}[
    every node/.style={font=\small},
    axis/.style={-{Stealth[length=6pt]}, thick},
    level/.style={thick, blue!70!black},
    ]

    \draw[axis] (0, 0) -- (12.5, 0) node[right] {$E$};
    \node[below] at (0, 0) {$0$};

    \foreach \cx in {2.5, 6.0, 9.5}{
      \draw[level, thick] (\cx, -0.4) -- (\cx, 0.4);
      \foreach \dx in {0, 0.1, 0.2, 0.3, 0.4}{
        \draw[level] (\cx+\dx, -0.25) -- (\cx+\dx, 0.25);
      }
    }
      \node[above=6pt] at (2.5 - .45, 0) {$\mathcal{E}_{0,0}$};
      \node[above=6pt] at (6.0 - .45, 0) {$\mathcal{E}_{1,0}$};
      \node[above=6pt] at (9.5 - .45, 0) {$\mathcal{E}_{2,0}$};

    \draw[decorate, decoration={brace, amplitude=5pt}]
    (2.5, 0.65) -- (2.5+.6, 0.65)
    node[midway, above=7pt] {$\order{1/R_{\text{cl}}^2}$};

    \draw[decorate, decoration={brace, amplitude=5pt, mirror}]
    (0.0, -0.65) -- (2.5, -0.65)
    node[midway, below=7pt] {$\order{1/\ell_{\delta}^2}$};

    \draw[decorate, decoration={brace, amplitude=5pt, mirror}]
    (2.5, -0.65) -- (6, -0.65)
    node[midway, below=7pt] {$\order{1/\ell_{\delta}^2}$};

    \draw[gray, thin, dashed] (0.0,       0.0)   -- (0.0,       -0.60);
    \draw[gray, thin, dashed] (2.5, -0.35)  -- (2.5,  -0.60);
    \draw[gray, thin, dashed] (6, -0.35)  -- (6,  -0.60);

  \end{tikzpicture}
  \caption{The low-energy spectrum for the probe fermion has clusters of states separated by \(\order{1/\ell_{\delta}^2}\), corresponding the discrete modes of the orthogonal matrix Airy problem, with a finer structure of eigenstates of the Laplacian on the orthogonal boundary sphere, with typical spacing of order \(\order{1/R_{\text{cl}}^2}\).}
  \label{fig:low-spectrum-sketch}
\end{figure}

Because of the confining harmonic trap, the spectrum of the full Hamiltonian \(H\) is discrete.
The spectrum of \(H_{\parallel}\) is the usual spectrum of a free particle on the sphere:
\begin{equation}
  \spec(H_{\parallel}) = \set{ \pm l(l+1),\quad l \in \setN_0} \, .
\end{equation}
The spectrum of \(H_\perp\) is also discrete (see also Appendix~\ref{sec:discr-spectr-h_perp}),
with eigenvalues of order \(\order*{1}\), since \(u\) is a dimensionless variable and all the parametric dependence is contained in the prefactor \(\left( {\mu}{\omega}^2 \right)^{1/3}\):
\begin{align}
	H_\perp \ket{\chi_n(u)} &= \chi_n \ket{\chi_n(u)}, & \braket{\chi_n|\chi_m} &= \delta_{n,m}.
\end{align}
All in all, at low energy, where we expect the spectrum of \(H\) to be that of the edge quasiparticles, the energy depends on
two numbers, \(n\) and \(l\), and has the form
\begin{equation}
  \mathcal{E}_{n,l} = \frac{1}{\ell_{\delta}^2} \chi_n + \frac{1}{2 M_{\parallel} R_{\text{cl}}^2} l (l + 1 ) \, ,
\end{equation}
The spectrum starts from \(\mathcal{E}_{0,0} > 0\), with \(l = 0\) states separated by intervals of order \(\order{1/\ell_{\delta}^2}\), and a finer structure of \( l \ge 1\) states with spacing of order \(\order{1/R_{\text{cl}}^2}\) as sketched in Figure~\ref{fig:low-spectrum-sketch}.

The decomposition of the Hamiltonian is reflected in the decomposition of the Nambu fermion as
\begin{equation}
  \ket{\Psi(t,u,\Omega)} = \sum_n \frac{1}{\sqrt{\ell_{\delta}}} \ket{\chi_n(u)} \otimes \ket{\psi_n(t,\Omega)}.
\end{equation}
Looking for the lowest states of $H$ selects the lowest mode $\ket{\chi_0}$:
\begin{equation}
	\ket{\Psi_0(t,u,\Omega)} = \frac{1}{\sqrt{\ell_{\delta}}}\ket{\chi_0(u)} \otimes \ket{\psi_0(t,\Omega)},
\end{equation}
effectively separating the perpendicular from the parallel dependence.
The action evaluated on \(\Psi_0\) then gives
\begin{equation}
  \label{eq:probe-action}
	\begin{aligned}
		S[\psi_0] &= \int \dd{t} \dd{x} \Psi_0^\dagger(t, x) \pqty*{i\del_t{} + H} \Psi_0(t,x)\\
        &\approx R_{\text{cl}}^2   \int \dd{t} \dd{u} \dd{\Omega} \psi_0^*(r, \Omega) \chi_0^\dagger(u) \pqty*{ i\del_t{} + \frac{1}{\ell_{\delta}^2} H_{\perp} + \frac{1}{2 R^2_{\text{cl}}} H_{\parallel} } \chi_0(u) \psi_0(t, \Omega)\\
                & =  R_{\text{cl}}^2  \int \dd{t} \dd{\Omega} \psi_0^*(t, \Omega) \pqty*{ i\del_t{} + \mathcal{E}_{0,0} + \frac{1}{2 R_{\text{cl}}^2 M_{\parallel}} \Laplacian_{S^2}} \psi_0 (t, \Omega),
	\end{aligned}
\end{equation}	
where
\begin{subequations}%
  \label{eq:probe-parameters}
  \begin{align}
    \mathcal{E}_{0,0} &= \frac{ \chi_0}{\ell_{\delta}^2}   \, , \\
    \frac{1}{M_{\parallel}}
        &= {\braket{\chi_0 | \boldsymbol{\tau}_{3} \chi_0}} \,.
  \end{align}
\end{subequations}
We get the action of a fermion localized on the surface \(r = R_{\text{cl}}\) with a ground state energy
\begin{equation}
  \mathcal{E}_{0,0} = \order*{\omega \left( \frac{2\mu}{\omega} \right)^{1/3}}
\end{equation}
and an effective mass \(M_{\parallel} = \order{1}\).
As anticipated in the introduction, we can identify this energy with the dominant contribution to the odd-even splitting \(\Delta_{\text{stag}}(Q)\):
\begin{equation}
   \Delta_{\text{stag}}(Q) = \mathcal{E}_{0,0}(Q) + \dots  
\end{equation}
It follows that in this approximation the universal coefficient \(\chi \) in Eq.~\eqref{eq:Son-formula} is the energy of the lowest eigenstate of \(H_{\perp}\):
\begin{equation}
  \chi = \chi_0 \, .  
\end{equation}

The quantities $\mathcal{E}_{0,0}$ and $M_{\parallel}$ are, in the probe approximation, not new parameters, but result from the diagonalization of \(H_{\perp}\).
One general observation is that \(M_{\parallel}\) depends on the particle-hole mixing at the edge (where the gap is minimal and the mixing is expected to be large), and by construction
\begin{equation}
  \abs{M_{\parallel}} \ge 1 \, .
\end{equation}
Unfortunately, this diagonalization cannot be performed with elementary methods but would require numerical methods.
As we undertake a more general numerical study in Section~\ref{sec:numerics}, we refrain from doing this here and refer to Table~\ref{tab:bdg-fit-parameters}.

Beyond reproducing the leading scaling argued in~\cite{Son2007}, the probe approximation also gives the explicit form of the fermion spectrum.
A natural concern is that the \ac{lda} breaks down precisely at the edge, which is where the physics of interest is localized.
However, the associated error is under systematic control: viewing the \ac{lda} gap profile as the leading term in the large-charge gradient expansion of Eq.~\eqref{eq:sigma-gradient-expansion}, corrections are organized in inverse powers of \(\mu\).
The leading correction to the ground-state energy takes the form
\begin{equation}
  \mathcal{E}_{0,0} = \frac{ \chi_0}{\ell_{\delta}^2} \pqty*{1 + \order*{\pqty*{\frac{\omega}{2 \mu}}^{2/3}}}  \, .
\end{equation}
So the results of this section are the leading terms in a controlled \(1/\mu \) expansion, which justifies the sharp-edge approximation in Eq.~(\ref{eq:sharp-edge}) as the leading term in a controlled expansion.
The exponent \(2/3\) can be obtained from the expression in Eq.~\eqref{eq:sigma-gradient-expansion}, which shifts the edge by an amount of order \(\order{\ell_{\delta}}\).
However a more systematic analysis requires going beyond the \ac{bdg} treatment, which motivates the \ac{eft} construction of the following section.

\section{EFT}\label{sec:eft}

In this section, we want to go beyond the probe limit, and describe the unpaired fermion within a large-charge \ac{eft} that does not depend on large-\(N\) or mean-field arguments.
This requires a generalization of the constructions in the literature to an odd number of fermions and allows us to compute the scaling laws and describe the interactions between the bulk superfluid and the edge fermion.
The price to pay is that, while we can compute the universal scaling laws, it will be necessary to add new \acp{lec} that, as usual, are parameters of the \ac{eft} and will have to be computed independently.
We still start again from the assumption that the fermion is localized on the cloud edge.

\paragraph{The superfluid EFT.}
Our starting point is the \ac{eft} of Son and Wingate~\cite{Son:2005rv} for the bulk, supplemented with the bosonic edge terms found by Hellerman and Swanson~\cite{Hellerman:2020eff}. 
The \ac{eft} in the bulk is written in terms of
\begin{itemize}
\item the Schrödinger-invariant combination
  \begin{equation}
    X = \dot{\theta} - \frac{1}{2} (\nabla \theta)^2 - V(r),
  \end{equation}
  where \(\theta\) is the field that realizes the \(U(1)\) symmetry non-linearly and is to be expanded around \(  \braket{\theta} = \mu t\),
\item its space derivatives \(\nabla X\),
\item the Schrödinger-invariant operator
  \begin{equation}
    Z = \Laplacian V - \frac{1}{3} (\Laplacian \theta)^2 \, .
  \end{equation}
\end{itemize}
For each operator in the \ac{eft}, it is necessary to take into account both the scaling dimension and the \(\mu\)-scaling \(\mudim{\Op}\), \emph{i.e.} the parametric scaling of its contribution to the ground state energy (see Table~\ref{tab:scalings}).
In the large-charge expansion (or, equivalently, the large-\(\mu\) limit), this is the right counting parameter.

\begin{table}
  \centering
  \begin{tabular}{lccc}
    \toprule
    &\(X\)& \(\nabla X\)& \(Z\) \\
    \midrule
    bulk & \(8/5\) & \(7/5 \) & \( 6/5\)\\
    edge && \(5/7\) & \(2/7\)\\
    \bottomrule
  \end{tabular}
  \caption{\(\mu\)-scaling for the EFT operators. The edge dimensions are understood for the operator multiplied by \(\delta(X)\).}
  \label{tab:scalings}
\end{table}

The leading-in-\(\mu\) operator in the bulk is \(X\), so we can write the bulk \ac{eft} Lagrangian as an integer expansion in \(\nabla X\) and \(Z\):
\begin{equation}
  L_{\text{bulk}} = c_0 X^{5/2} \sum_{m,n} c_{m,n} \frac{\abs{\nabla X }^{2m} Z^n}{X^{3m + 2n}} ,
\end{equation}
or, equivalently, the generic bulk operator has the form~\cite{Hellerman:2020eff}
\begin{equation}
  \Op_{\text{bulk}}^{(m,n)} = \abs{\nabla X }^{2m} Z^n X^{5/2 - (3m + 2n)}
\end{equation}
and has \(\mu\)-scaling
\begin{equation}
  \mudim{\Op^{(m,n)}_{\text{bulk}}{}}  = 4 - 2 (m + n) \, .
\end{equation}
The edge is defined by the condition \(X = 0\), so the action can only depend on \(\nabla X\) and \(Z\).
Now the leading operator is \(\nabla X\) and the most generic operator that can appear in the \ac{eft} Lagrangian is
\begin{equation}
  \tilde{\mathcal{Z}}^{(p)} = \delta(X) Z^p \abs{\nabla X}^{1 + 4/3(1 - p)}  
\end{equation}
and has \(\mu\)-scaling
\begin{equation}
  \mudim{\tilde{\mathcal{Z}}^{(p)}} = \frac{5 - 2 p}{3} \, .
\end{equation}
The operator \(\tilde{\mathcal{Z}}^{(2)}\) gives a contribution \(\order{\mu^{1/3}}\) to the ground-state energy, which is the same order as the one expected from our analysis in Section~\ref{sec:probe-approximation}.
This term is part of the superfluid \ac{eft} of the condensed Cooper pairs and should not be confused with the effect of the unpaired edge fermion, which appears when the total number of particles is odd.

The leading bulk and edge operators are, respectively, \(\Op_{\text{bulk}}^{(0,0)}\) and \(\tilde{\mathcal{Z}}^{(0)}\), and from here on we will use the action
\begin{equation}
  S_{\text{bulk}} = \int \dd{t} \dd^3{x} \bqty*{ c_0\, X^{5/2} + \dots + \delta(X) \pqty*{ d_0 \abs{\nabla X}^{7/3}+\dots} } \ .
\end{equation}

\paragraph{The co-area theorem and the fermion at the edge.}
We want to supplement the \ac{eft} action with the contribution of one extra fermion that, as we said above, should be localized at the edge.
Based on the ingredients that we have, the most generic two-fermion term at the edge has the form
\begin{equation}
  S_{\tilde \Psi} = \int \dd{t} \dd^3{x} \delta(X) \tilde \Psi^{\dagger} \tilde{\mathcal{K}}( \del_t{}, \Laplacian{}, \nabla X, Z) \tilde \Psi ,
\end{equation}
where \(\tilde{\mathcal{K}}\) has scaling dimension two, and should be thought of as an expansion in \(1/ \mu\).

The field-valued \(\delta\) function is a bit unusual and would have to be regulated to properly describe the phonon fluctuations, as already pointed out in the original paper~\cite{Hellerman:2020eff}. 
Instead we choose to rewrite the edge action in an equivalent form via the co-area theorem.
Given a (sufficiently well-behaved) function $X : \setR^4 \to \setR$,
one can rewrite integrals over \(\setR^4\) as a decomposition over the levels \(X=s\):
\begin{equation}
  \int \dd t \dd^3{x} f(t,x) = \int_{\setR} \dd{s} \int_{\Sigma_s = X^{-1}(s)} \dd t\dd{\Sigma_s} 
  \frac{f(t,s,\Sigma)}{\abs{\nabla X}},
\end{equation}
where \(\dd{\Sigma_s}\) is the volume element over the level set of $X$, $\Sigma_s = \set{X=s}$.
Effectively, we are slicing the integral over surfaces at constant particle density, as $\rho(t, x) \propto (X(t, x))^{3/2}$.

The edge terms contain a \(\delta\) function, which turns into $\delta(s)$ and localizes the first integral.
Then we find for the leading \(\tilde{\mathcal{Z}}^{{0}}\) operator
\begin{equation}
  \int \dd{t} \dd^3{x} \delta(X) (\nabla X)^{7/3} = \int \dd{t} \int_{X^{-1}(0)} \dd{\Sigma_0} \abs{\nabla X}^{4/3}.
\end{equation}
We now have two equivalent descriptions: one as a bulk action, the other in terms of a dynamical surface \(\Sigma_0 = X^{-1}(0)\).

Higher-order terms in the edge \ac{eft} can be treated in the same way and the general insertion takes the form
\begin{equation}
  \mathcal{Z}^{(p)} = \eval*{Z^p \abs{\nabla X}^{4/3(1 - p)}  }_{\Sigma_0} \, .
\end{equation}
Using the co-area theorem we can also rewrite the fermionic contribution. After a field redefinition it becomes
\begin{equation}
  S_{\Psi} = \int \dd{t} \dd{\Sigma_0} \Psi^{\dagger}(t, \Sigma_0) \mathcal{K} \Psi(t, \Sigma_0) ,
\end{equation}
where
\begin{align}
  \Psi(t, \Sigma_0) &= \eval*{\abs{\nabla X}^{1/2} \tilde{\Psi}(t, x)}_{\Sigma_0} ,\\
  \mathcal{K} &= \eval*{\abs{\nabla X}^{1/2} \tilde{\mathcal{K}} \abs{\nabla X}^{-1/2}}_{\Sigma_0} \, .
\end{align}
Expanding the operator \(\mathcal{K}\) at leading order in the derivatives gives
\begin{equation}
  S_\Psi = \int \dd{t} \dd{\Sigma_0}  \bqty*{\Psi^\dagger \pqty*{ i \del_t{} - V + \frac{1}{2M_{\parallel}} \Laplacian{} - \mathcal{E}( \abs{\nabla X}, Z )}  \Psi  } .
\end{equation}
The leading-in-\(\mu\) term in the coupling to the bosonic operator \(\abs{\nabla X}\) is fixed by scale invariance:
\begin{equation}
  \mathcal{E}(\abs{\nabla X}) = e_0 (\abs{\nabla X})^{2/3} + \dots \, ,
\end{equation}
where $e_0$ is a dimensionless parameter.
The exponent of the leading operator is entirely fixed by dimensional analysis since there are no intrinsic scales at unitarity.

The physical fermion \(\Psi \) has unit charge under the \(U(1)\) symmetry that is broken in the bulk.
The standard method for writing an \ac{eft} in which the fields only transform under the unbroken symmetries~\cite{Coleman:1969sm,Callan:1969sn} is to factor out the condensate phase and introduce a new, neutral, fermion \(\psi\):
\begin{equation}
  \Psi(t, \Sigma_0)  =  \eval*{e^{-i \theta(t, x) }}_{\Sigma_0} \psi(t, \Sigma_0) .
\end{equation}
Rewriting the action in terms of the uncharged field \(\psi\), we find%
\begin{multline}
  S_{\psi} = \int \dd{t}\dd{\Sigma_0} \Bigg[ i \psi^{\dagger} \del_t \psi - \frac{1}{2M_{\parallel}} \abs{\nabla \psi}^2  + e_0  \abs{\nabla X}^{2/3} \psi^\dagger \psi  \\
  + \frac{1}{2} \pqty*{1 - \frac{1}{M_{\parallel}}} (\nabla \theta)^2 \psi^{\dagger} \psi  - \frac{i}{2M_{\parallel}} \nabla \theta \cdot \psi^\dagger \overleftrightarrow{\nabla} \psi  \Bigg] \, ,
\end{multline}
where we have introduced the fermionic current
\begin{equation}
  j = \psi^\dagger \overleftrightarrow{\nabla} \psi = \psi^{\dagger} \nabla \psi - \nabla \psi^{\dagger} \psi \, .
\end{equation}
This form of the action extends the large-charge superfluid \ac{eft} to include an unpaired fermion at the boundary coupled to the superfluid in the bulk.
It is useful to distinguish two physical effects.
\begin{itemize}
\item The first is the fact that the quasiparticle effective mass \(M_{\parallel}\) is different from the mass \(M = 1\) of the fundamental fermions in the bulk, and the corresponding coefficient \(1 - 1/M_{\parallel}\) controls the \((\nabla \theta)^2 \psi^{\dagger} \psi\) interaction.
  In the probe approximation we have seen that \(M_{\parallel} > 1\), so the coefficient of this interaction is positive (the fermion localizes where the Goldstone flow is minimal).
  It would be interesting if this coud be argued from \ac{eft} consistency arguments alone.  
\item The second is that the fermionic edge current is naturally coupled to the velocity of the Goldstone field \(\nabla \theta\), which acts as a dynamical medium in which the quasiparticle propagates.

\end{itemize}

\paragraph{Action at the superfluid saddle (or, the probe approximation).}

The first thing that we can do with our \ac{eft} is to reproduce the results of the probe approximation.
Neglecting the backreaction of the edge fermion is the same as freezing the bulk field \(\theta(t, x) \) to its saddle \(\theta = \mu t\) and keeping only the \ac{vev} for the bosonic terms in the fermion action.

The field \(\eval*{\nabla X}_{\Sigma_0}\) has scaling dimension three and its expectation value on the semiclassical ground state is%
\footnote{This expectation value coincides with the value of the ``electric field'' \(\mathcal{E}\) in~\cite{Son2007}.}
\begin{equation}
  \braket{\eval*{\nabla X}_{\Sigma_0}} =  \omega^2 R_{\text{cl}} = \omega \sqrt{2\mu} \, .
\end{equation}
Perhaps unsurprisingly, as we are describing the physics at the edge, this expectation value is entirely fixed by the scale $\ell_\delta$ that we had found in the previous section:
\begin{equation}
  \braket{\eval*{\nabla X}_{\Sigma_0}} = \ell_\delta^{-3} \, .
\end{equation}
With this, we can evaluate the fermionic action at the superfluid saddle, where the bosonic field is \(\theta = \mu t\) and \(\Sigma_0\) is a sphere of radius \(R_{\text{cl}}\):
\begin{equation}
  \braket{S_{\psi}} = R_{\text{cl}}^2 \int \dd{t}\dd{\Omega} \bqty*{ i \psi^{\dagger} \del_t \psi + e_0  (\omega \sqrt{2 \mu})^{2/3} \psi^\dagger \psi  + \frac{1}{2 M_{\parallel} R_{\text{cl}}^2} \psi \Laplacian_{S^2} \psi } \, .
\end{equation}
This is precisely the same form that we had obtained in the probe approximation in Eq.~(\ref{eq:probe-action}), with the identification
\begin{equation}
  \mathcal{E}_{0,0} =  e_0  (2 \mu \omega^2)^{1/3} \, .
\end{equation}
The parametric dependence of \(\mathcal{E}_0\) and \(M_{\parallel}\) on \(\mu\) and \(\omega\) is not fixed by scale invariance alone, but we still recover the same result as in Eq.~(\ref{eq:probe-parameters}).
This is a nice selfconsistency check of our construction.

The universal coefficient \(\chi\) is identified with the low-energy coefficient \(e_0\).
As we have pointed out already, now we have two dimensionless couplings \(e_0\) and \(M_{\parallel}\) that are parameters of the \ac{eft} and have to be computed independently and, as we will see in Section~\ref{sec:numerics}, can be extracted from a numerical analysis in the probe approximation.
However, the \ac{eft} construction has the advantage that it does not depend on mean-field-type assumptions, and can be readily generalized.

\paragraph{Higher terms in the \(1/Q\) expansion.}
One possible generalization is to take into account the higher terms in the \(1/Q\) expansion, which go beyond the \ac{lda} and describe the effects related to breaking of translation invariance due to the harmonic trap (see~\cite{Hellerman:2023myh} for a discussion).

The operator \(\mathcal{E}(X) \Psi^{\dagger} \Psi\) that we have written is actually just the first term of an expansion in \(\nabla X\) and \(Z\):
\begin{equation}
  \mathcal{E}(\nabla X, Z) \Psi^{\dagger} \Psi = \sum_{p=0} e_p \abs{ \nabla X}^{(2 - 4 p)/3} Z^p \Psi^{\dagger} \Psi  ,
\end{equation}
and its contribution to the edge fermion grounds state energy is
\begin{equation}
  \braket{\mathcal{E}(\nabla X, Z)} \Psi^{\dagger} \Psi = \sum_{p=0} e_p \pqty*{ 2 \mu \omega^2 }^{(1 - 2 p)/3} \omega^{2p} \Psi^{\dagger} \Psi , 
\end{equation}
so we find an expansion in powers of \(1/\mu^{2/3}\) (or, equivalently, in \(1/Q^{2/9}\)):
\begin{equation}
  \mathcal{E}_{0,0} = e_0 \pqty*{2 \mu \omega^2}^{1/3} \pqty*{ 1 + \frac{e_1 \omega^2}{e_0 \pqty*{2 \mu \omega^2}^{2/3}} + \dots}  .
\end{equation}
While in the \ac{lda}, we could only find the leading $\mu$ scaling of the couplings, we are now able to compute corrections in terms of the \ac{lec}s of the higher-order terms in the \ac{eft} Lagrangian.

\paragraph{Phonons.}

When considering the phonons, we go beyond the probe approximation, so we are able to see effects here which are not captured in the \ac{bdg} treatment of the previous section or elsewhere in the literature.
The edge/bulk system is coupled via fermion-Goldstone interactions that we will find in the following and which are lost when the backreaction of the condensate is frozen out.
Given the way in which we have formulated the edge action, there will be both contributions from the expansion of the Lagrangian density and from the dynamics of the surface over which the density is integrated.

We start from the expansion \(\theta = \mu t + \pi\), and expand both the field in the integrand and the geometry of the edge.
For this purpose it is convenient to introduce a bookkeeping field \(s\) that parametrizes the fluctuations of the edge around \(R_{\text{cl}}\).
The surface \(\Sigma_0\) is then defined by the condition
\begin{equation}
  r(t, \Omega) = R_{\text{cl}} + s(t, \Omega) \, .
\end{equation}
The field \(s \) is not independent, and its value is fixed by imposing the constraint
\begin{equation}
  X(t, R_{\text{cl}} + s, \Omega)   = 0 \, .
\end{equation}
Expanding at first order in \(\pi\) and \(s\), we find
\begin{equation}
  X(t, R_{\text{cl}}+ s, \Omega)  = X_0(t, R_{\text{cl}} + s, \Omega) + \dot \pi = X_0(R_{\text{cl}}) + s \eval*{\pdv{X}{r}}_{R_{\text{cl}}} + \dot \pi,
\end{equation}
where \(X_0(R_{\text{cl}}) = \eval*{\mu - \omega^2/2 r^2}_{R_{\text{cl}}} = 0\).
From this we find the relation
\begin{equation}
  s = - \frac{\dot \pi}{X_0'(R_{\text{cl}})} = \frac{\dot \pi}{ \omega \sqrt{2\mu}} = \ell_{\delta}^3 \dot \pi \, .
\end{equation}
Using \(s\) we can expand all the geometric objects in the action.
For the measure, we find
\begin{equation}
  \dd{\Sigma_0} =  R_{\text{cl}}^2 \pqty*{1 + \frac{2s}{R_{\text{cl}}}} \dd{\Omega},
\end{equation}
where \(\dd{\Omega}\) is the standard measure on the unit sphere, while for the Laplacian,
\begin{equation}
  \Laplacian_{\Sigma_0} = \frac{1}{\pqty*{R_{\text{cl}}+ s}^2} \Laplacian_{S^2} = \frac{1}{R_{\text{cl}}^2} \pqty*{1 - \frac{2s}{R_{\text{cl}}}} \Laplacian_{S^2} \, ,
\end{equation}
so that at leading order \(\dd{\Sigma_0} \psi^{\dagger} \Laplacian_{\Sigma_0}{\psi}\) remains invariant.
As for the integrand, it is necessary to expand \(\abs{\nabla X}^{2/3}\), which gets two contributions, one from the expansion of \(\theta\), the other from the change in the radius:
\begin{equation}
  \abs{\nabla X}^{2/3} = \pqty*{\omega \sqrt{2\mu}}^{2/3} - \frac{2}{3 \pqty*{\omega \sqrt{2\mu}}^{1/3}} \pqty{\omega^2 s - \del_r \dot \pi} .
\end{equation}
Putting everything together we get the leading fermion-fermion-Goldstone interaction:
\begin{equation}
  S_{\psi \psi \pi} = R_{\text{cl}}^2 \int \dd{t} \dd{\Omega} \bqty*{ \frac{2 \ell_{\delta}}{3} \pqty*{\frac{4}{R_{\text{cl}}} \dot \pi - \del_r \dot \pi} \psi^{\dagger} \psi +  \frac{i \dot \pi}{\mu}  \psi^{\dagger} \del_t \psi - \frac{i}{2M_{\parallel}} \nabla \pi \cdot \psi^\dagger \overleftrightarrow{\nabla} \psi }\, .
\end{equation}
The term coupling to the current has a natural interpretation. %
The derivative of the Goldstone boson \(\nabla \pi\) describes the tangential superflow.
It acts as a background connection for the fermion that is coupled to it via the conserved edge current.
In other words, the edge excitations propagate in a medium that has a local superflow and their motion is shifted according to whether they travel with or against that flow (Doppler effect).
A similar interpretation can be given also for the coupling of \(\dot \pi\) to the time current.

\medskip

The \ac{eft} methodology adopted in this section goes beyond the \ac{bdg} method in the probe limit used in Section~\ref{sec:probe-approximation}.
Apart from reproducing the same prediction for the scaling of the relevant observables, it allows us to understand the subleading corrections and, more importantly from a conceptual point of view, it allows us to capture the interaction between the unpaired fermion and the bulk bosons consisting of Cooper pairs.

\section{Numerical analysis}
\label{sec:numerics}

In this section we report on the results of a numerical study.
We work in the probe approximation with a sharp-edge \ac{lda} gap profile, using the parameter \(y\)
from large-\(N\) or \ac{qmc} studies as in Section~\ref{sec:probe-approximation}.
Although the \ac{lda} breaks down precisely at the edge, the associated error is systematically controlled, and the results below are the leading terms in an expansion in inverse powers of  \((2 \mu / \omega)^{2/3}\).

In constrast with our treatment in Section~\ref{sec:probe-approximation}, we will not make any assumptions on the spectrum or on the localization of the low-lying states here.
Instead, we will measure these quantities with numerical methods to obtain an independent confirmation of our results.

The idea is to discretize the \ac{bdg} Hamiltonian in Eq.~(\ref{eq:BdG-Hamiltonian}) and look for the lowest eigenstates,
\begin{equation}
  H \begin{pmatrix}
    u \\ v
  \end{pmatrix} = E \begin{pmatrix}
    u \\ v
  \end{pmatrix} \, .
\end{equation}
The system is confined by the harmonic oscillator potential \(V = \omega^2/2 r^2\) and the spectrum is discrete.
Making use of the \(SO(3)\) symmetry we can expand the quasiparticle amplitudes in a basis of spherical harmonics,
\begin{align}
  u_n(\mathbf{r})
  &=
  \sum_{l=0}^{\infty}\sum_{m=-l}^{l}
  \frac{u_{nl}(r)}{r}Y_{lm}(\Omega),
  &
  v_n(\mathbf{r})
  &=
  \sum_{l=0}^{\infty}\sum_{m=-l}^{l}
  \frac{v_{nl}(r)}{r}Y_{lm}(\Omega),
\end{align}
and label the states using three quantum numbers \(n, l, m\).
As long as the \(SO(3)\) symmetry is not broken, energies will only depend on \(n\) and \(l\) and states will be \(2l + 1\) degenerate.

The decomposition into spherical harmonics corresponds to a decomposition of the Hamiltonian into a radial and an angular part:
\begin{align}
  H &= H_0 - \frac{1}{2 r^2} l (l + 1) \boldsymbol{\tau}_3 ,\\
  H_0 &= h_0 \boldsymbol{\tau}_3 + \sigma(r) \boldsymbol{\tau}_1, \\
  h_0 &= -\frac{1}{2} \frac{\dd^2{}}{\dd{r^2}} - \mu + V(r).
\end{align}
For the gap \(\sigma(r)\), we are in the \ac{lda}, which means that
\begin{equation}
  \sigma(r) = y \pqty*{\mu - V(r)},  
\end{equation}
where the coefficient \(y\) can be computed either in a large-\(N\) approximation or via \ac{qmc} simulations (see Table~\ref{tab:gap-mu-y}).
In the following, we will concentrate on the \(l = 0\) sector, since we are interested in the low-lying spectrum of the Hamiltonian.

The reduced radial functions \(u_{n,0}(r)\) and \(v_{n,0}(r)\) are placed on an interval \(r \in [ 0, r_{\text{max}}]\) with Dirichlet boundary conditions to avoid singular behavior at the origin.
The size of the interval is
\begin{equation}
  r_{\text{max}} = R_{\text{cl}} + \frac{b}{\sqrt{\omega}} , 
\end{equation}
where $b$ is a dimensionless padding parameter, typically of order
$10$--$12$. This ensures that the classically allowed superfluid
region and the edge-localized quasiparticle wavefunction are well
contained inside the numerical interval.

The discretization is performed on a uniform grid of spacing \(\dd{r} = r_{\text{max}}/(N_r + 1)\), with nodes at positions \(r_i = i \dd{r}\).
The resulting real symmetric \(2N_r \times 2 N_r\) matrix is projected on the Krylov subspace generated by
\begin{equation}
  \mathcal K_m(H_0,q)
  =
  \Sp\{q,H_0q,H_0^2q,\ldots,H_0^{m-1}q \},
\end{equation}
where \(q \) is a trial vector.
This targets the near-zero \ac{bdg} spectrum while avoiding the ultraviolet modes associated with the grid cutoff~\cite{Arnoldi:1951dpg}.

\begin{figure}[t]
  \centering
  \includegraphics[width=.75\textwidth]{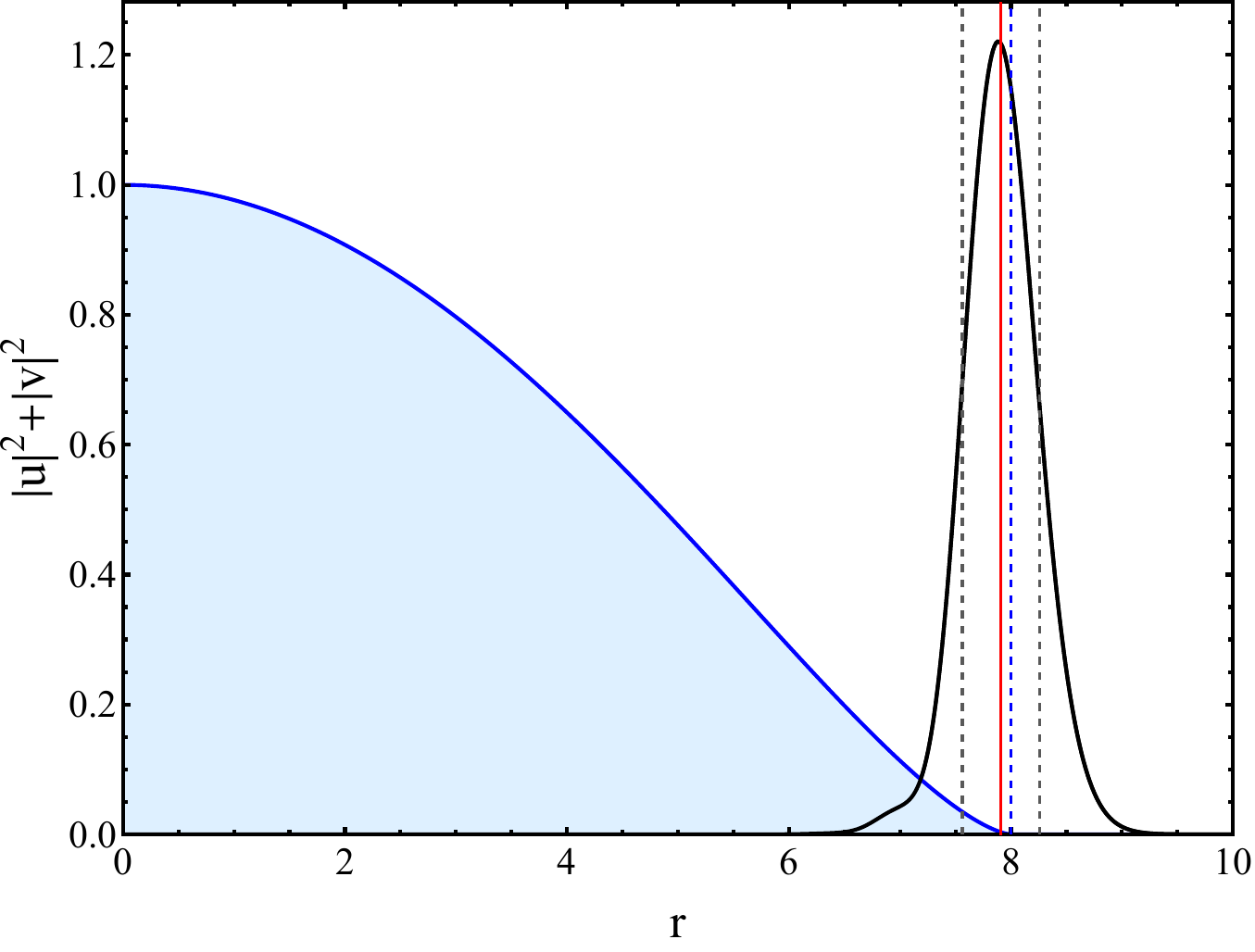}
  \caption{Radial probability density for the lowest eigenstate for \(y = 1.22\) and \(\mu = 32 \omega\), corresponding to \(R_{\text{cl}} = 8/\sqrt{\omega}\) (blue dotted line), $\bar{r}=0.79$ (red line) and with variance $w=0.35$ (gray dashed lines).
  The quasiparticle is clearly localized around the edge of the droplet. Superposed and shaded is the local bulk density, normalized to unity.}
  \label{fig:quasiparticle-density}
\end{figure}

Given the \(n\)-th eigenstate of the radial Hamiltonian, the quasiparticle probability density is
\begin{equation}
  \rho_n(r) = \abs{u_n(r)}^2 + \abs{v_n(r)}^2 = \abs{u_{n,i}}^2 + \abs{v_{n,i}}^2,
\end{equation}
where \(u_{n,i} = u_n(r_i)\).
The center of the distribution and its variance are
\begin{align}
  \bar{r}_n &=  \braket{r}_n = \frac{\int \dd{r} r \rho_n(r)}{\int \dd{r} \rho_n(r)} = 
              \sum_i r_i\left( \abs{u_{n,i}}^2 + \abs{v_{n,i}}^2 \right), \\
  w_n^2 &=  \braket{ \pqty{r - \bar r_n}^2}_n = \frac{\int \dd{r} \pqty{ r - \bar{r}_n}^{2} \rho_n(r)}{\int \dd{r} \rho_n(r)} = 
          \sum_i \pqty{r_i - \bar r_n}^2 \left(|u_{n,i}|^2+|v_{n,i}|^2\right).
\end{align}
For an edge mode, one expects that its center is on the cloud edge and its width is of the order of the size of the \(\delta \) layer (see Figure~\ref{fig:quasiparticle-density}):
\begin{align}
  \bar{r}_n &\approx R_{\text{cl}}  = \frac{\sqrt{2\mu}}{\omega} \, , & w_n &= \order{\ell_{\delta}} = \order{(\omega \sqrt{2\mu})^{-1/3} }.
\end{align}

In order to compute the effective mass without diagonalizing the full Hamiltonian \(H\), we treat the centrifugal term \(1/(2r^2) l (l + 1) \boldsymbol{\tau}_3\) as a perturbation with parameter \(\alpha = l (l + 1)\).
By the Hellmann--Feynman theorem, the variation of the energy is the expectation value of the variation of the Hamiltonian evaluated on the unperturbed state:
\begin{equation}
  S_n = \eval*{ \pdv{\mathcal{E}_{n,l}}{\alpha}}_{\alpha = 0} = \braket{ \pdv{H}{\alpha}}_n = \braket{\frac{\boldsymbol{\tau}_3}{2 r}}_n = \sum_i \frac{\abs{u_{n,i}}^2 - \abs{ v_{n,i}}^2}{2 r_i^2}.
\end{equation}
Defining the effective mass \(M_{\parallel}\) from
\begin{equation}
  \mathcal{E}_{n, l} = \mathcal{E}_{n,0} + \frac{\alpha}{2 R_{\text{cl}^2} M_{\parallel, n}}  + \dots ,
\end{equation}
we find immediately
\begin{equation}
  M_{\parallel, n} = \frac{1}{2 R_{\text{cl}}^2 S_n}  .
\end{equation}
The effective mass is numerically more delicate than the
energy and the width because the matrix element involves the difference
$\abs{u}^2 - \abs{v}^2$ rather than the positive density $\abs{u}^2 + \abs{v}^2$.
It is  therefore more sensitive to particle--hole cancellations and to
subleading corrections in the edge expansion (see also the expression in Eq.~\eqref{eq:probe-parameters}).

The computation was repeated for different values of \(\mu\), between \(\mu = 20 \omega\) and \(\mu = 700 \omega\) and several convergence checks were performed.
The number of grid points \(N_r\) was varied to ensure stability; the number of retained near-zero eigenvalues was varied to make sure that the same positive eigenvalues had been selected; the box padding was varied to ensure that the wavefunction was insensitive to the artificial bound at \(r = r_{\text{max}}\).

\begin{figure}[]
  \begin{center}
    \includegraphics[scale=0.57]{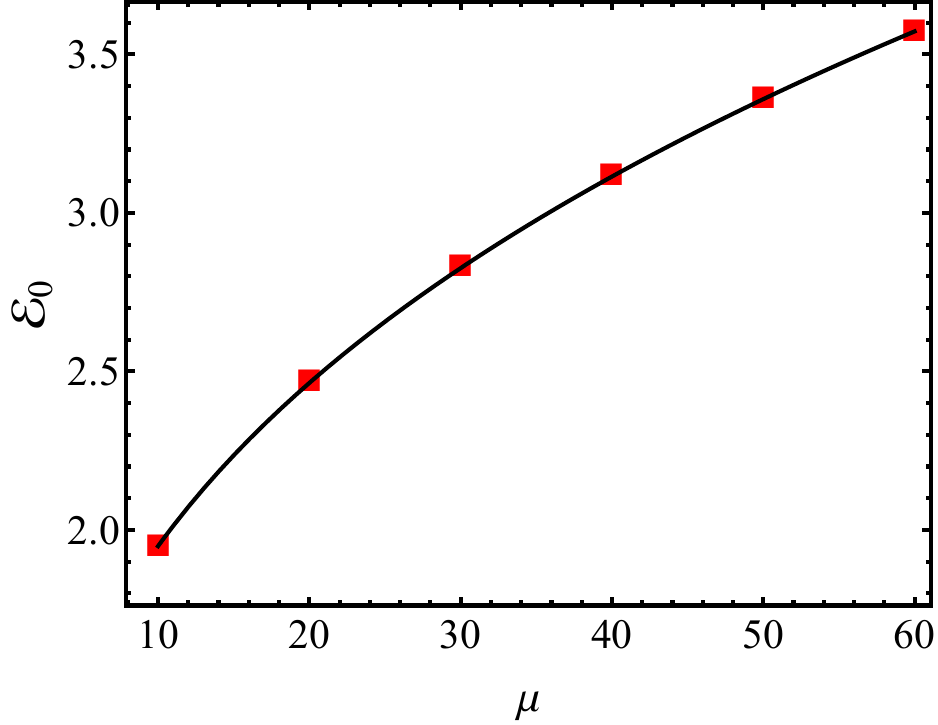}
    \includegraphics[scale=0.57]{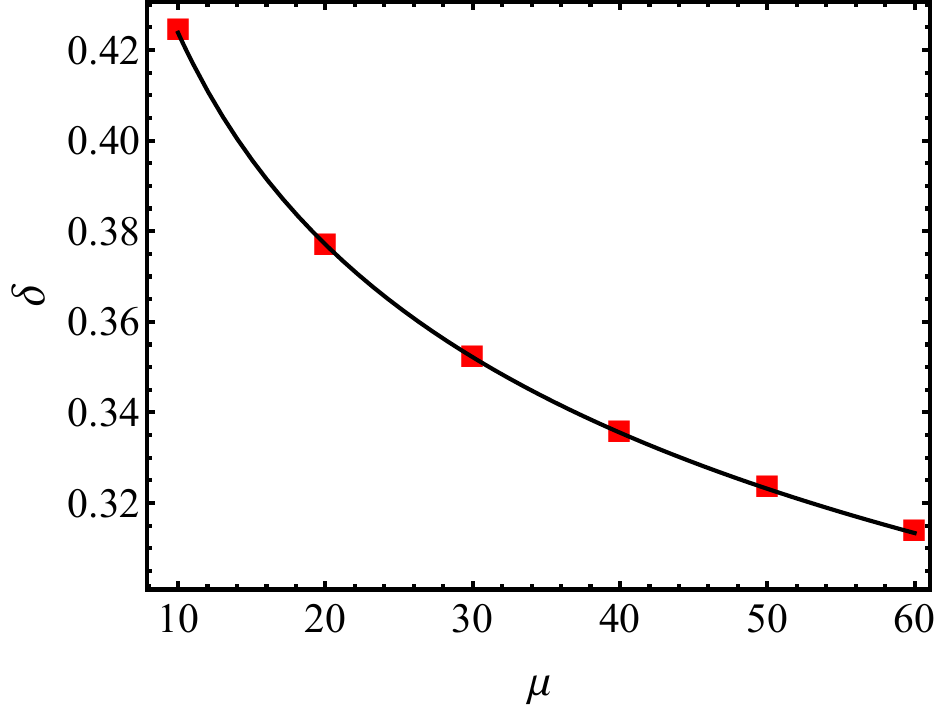}
    \includegraphics[scale=0.57]{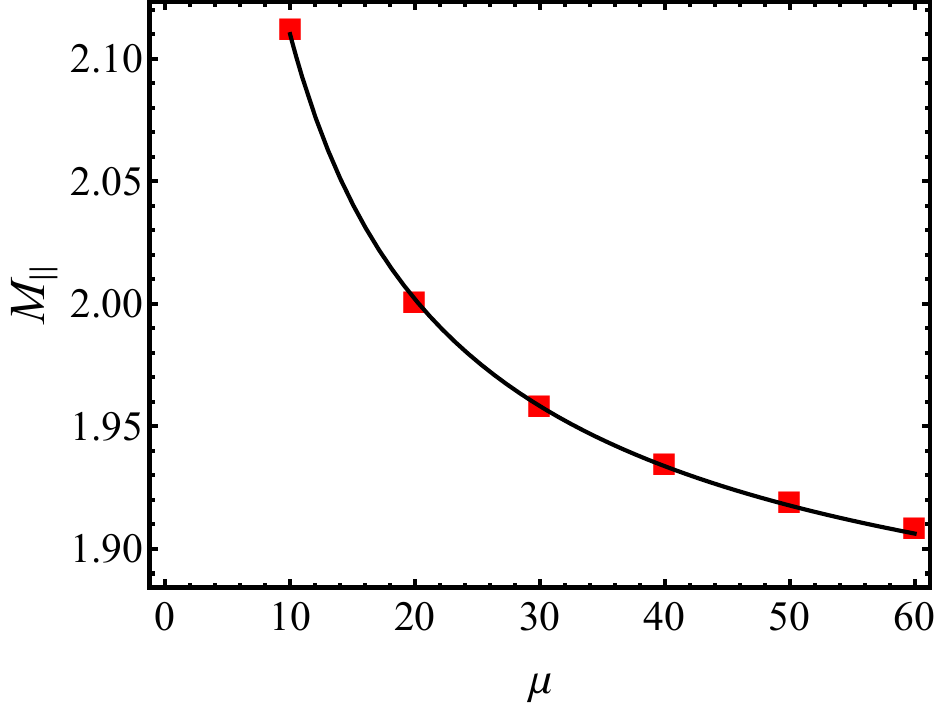}
    \caption{\small Plots of $\mathcal{E}_0$, $\delta$, $M_{\parallel}$ vs. $\mu$ for the QMC input data. The red squares are simulation data, generated as described in the text, and the solid black lines are best fits with parameters given in Table~\ref{tab:bdg-fit-parameters}. \label{fig:Allvsmu}}
\end{center}
\end{figure}

The numerical data was fit to dimensionless power laws, via a log-log power-law fit for the coupling of each observable \(\Op\) of the form
\begin{equation}
  C[\Op] \pqty*{\frac{2\mu}{\omega}}^{p[\Op]} \, .
\end{equation}
As the effective mass is sensitive to higher order corrections, we choose to fit
to the form
\begin{equation}
  \frac{1}{M_{\parallel}}
  =
  J_0+J_1\left(\frac{2\mu}{\omega}\right)^{-2/3}
  +\cdots .
\end{equation}
Example fits of $\mathcal{E}_0$, $\delta$, and $M_{\parallel}$ to simulation data are shown in Figure~\ref{fig:Allvsmu}.
\paragraph{Results.}

The calculation of the staggering energy takes as input the value of $y$, which measures the superfluid gap in units of the chemical potential.
Its large-\(N\) value~\cite{Veillette:2007zz}, as well as its value computed using \ac{qmc} simulations, is given in Table~\ref{tab:gap-mu-y}.
The uncertainty in $y$ arises from the uncertainties in the superfluid gap at unitarity~\cite{Carlson:2007omx} and the value of the Bertsch parameter~\cite{Carlson:2014pxa}. 

Our results are summarized in Table~\ref{tab:bdg-fit-parameters}.
The power-law exponents are remarkably close to the analytic predictions and all the coefficients \(C[\Op]\) are of order one, as one would expect from naturalness arguments in a strongly-coupled system without intrinsic scales.

Before comparing to existing simulation results, it is useful to separate clearly which ingredients of our analysis are taken as external input and which are genuine outputs.
The microscopic \ac{bdg} calculation uses as input the bulk equation-of-state parameter $\xi$ and the dimensionless homogeneous gap ratio $y=\Delta/\mu$, for which we consider both large-$N$ and \ac{qmc}-motivated values; these inputs are summarized in Table~\ref{tab:gap-mu-y}.
The probe-limit numerical diagonalization then yields the low-lying quasiparticle energies, their localization diagnostics, and the effective edge parameters such as $E_{0,0}$ and $M$, whose large-$Q$ scaling is compared to the analytic edge predictions in Table~\ref{tab:bdg-fit-parameters}.

\begin{figure}[t]
  \begin{center}
    \includegraphics[scale=0.485]{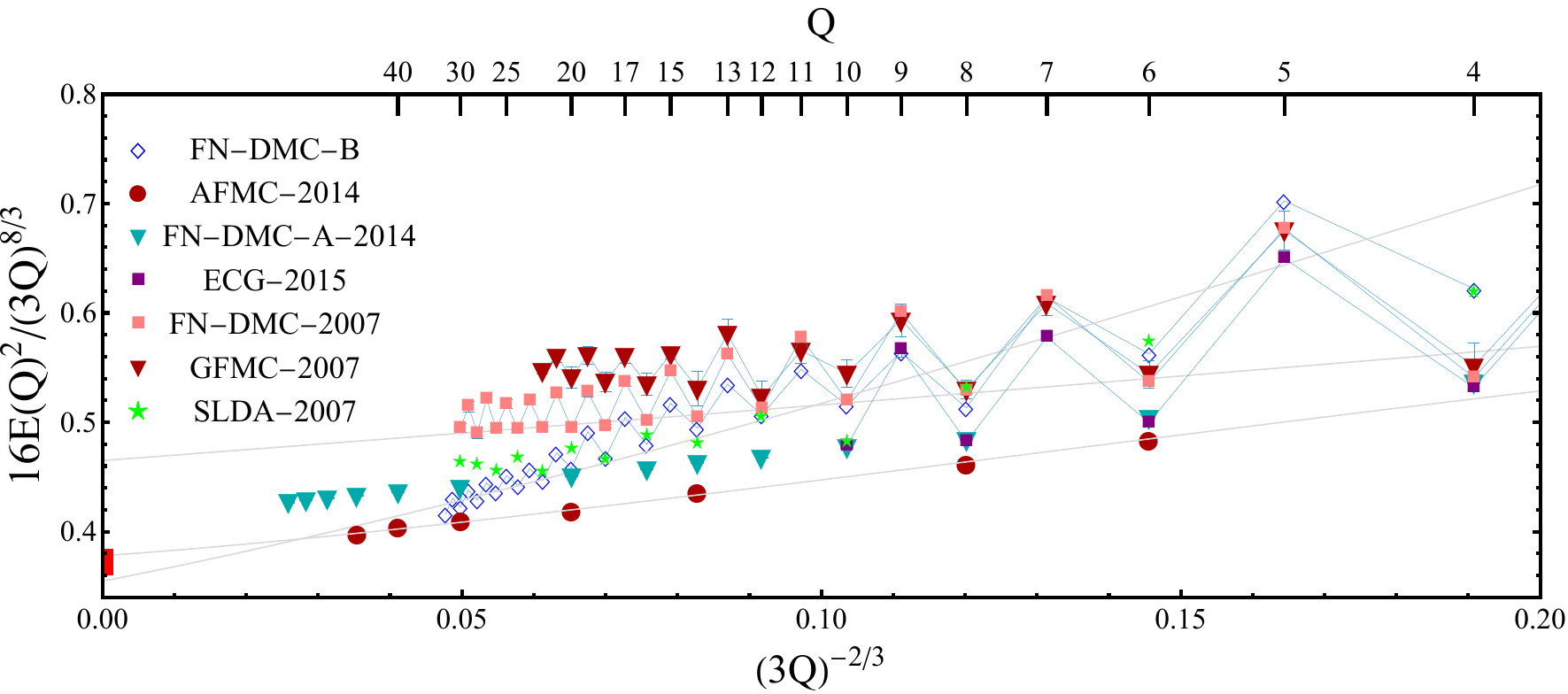}
    \caption{\small Plot of the energy $E(Q)$ of $Q$ fermions in a harmonic trap
      in a variable that asymptotes to the Bertsch parameter $\xi$ at
      large-$Q$. The (AFMC-2014, FN-DMC-2014, ECG-2015, FN-DMC-2007) data
      are taken from
      (\cite{Carlson:2014pxa,Yin_2015,Blume_2007}). The empty blue diamonds ---denoted FN-DMC-B--- are from Ref.~\cite{GandolfiPrivComm2026}. The green stars ---denoted SLDA-2007--- are density-functional theory data taken from Ref.~\cite{PhysRevA.76.040502}. The lines are best fits to the FN-DMC-B, AFMC-2014, and FN-DMC-2007 data, and the filled red rectangle is the value of the Bertsch parameter that is adopted in this work~\cite{Carlson:2007omx}.
       \label{fig:WernerALL}}
\end{center}
\end{figure}

The state of simulation data for the staggering energy for given $Q$ is given in 
Figure~\ref{fig:WernerALL}. The energy is plotted in a variable~\cite{Forbes:2012yp} that asymptotes to the Bertsch parameter $\xi$ at large-$Q$. The \ac{qmc} data (\textsc{afmc}-2014, \textsc{fn-dmc}-2014, \textsc{ecg}-2015, \textsc{fn-dmc}-2007) %
data are taken from (\cite{Carlson:2014pxa, Yin_2015, Blume_2007}). The empty blue diamonds ---denoted \textsc{fn-dmc}-B--- are from Ref.~\cite{GandolfiPrivComm2026}. 
The green stars ---denoted \textsc{slda}-2007--- are density-functional theory data taken from Ref.~\cite{PhysRevA.76.040502}.

\begin{figure}[t]
  \begin{center}
    \begin{tabular}{l}
      \includegraphics[scale=0.475]{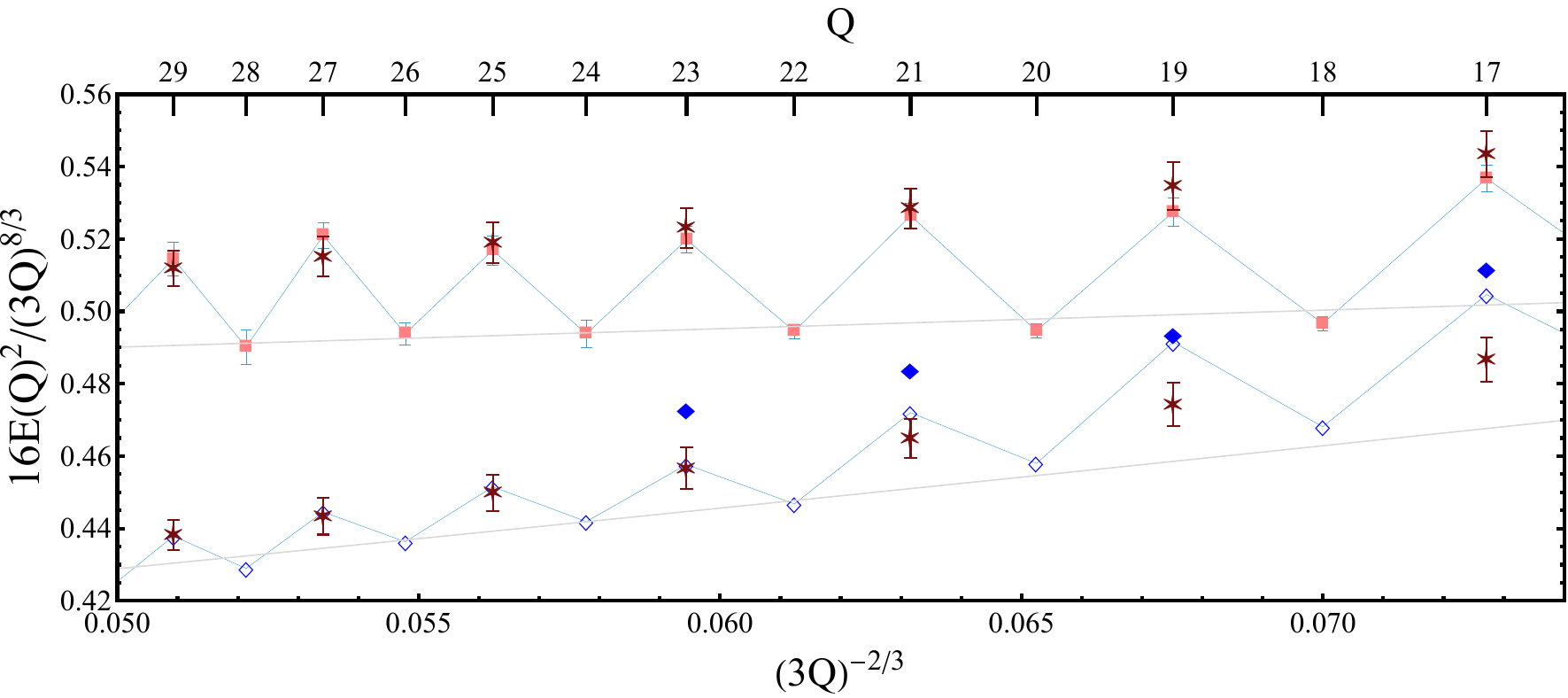} \\[2em]
      \includegraphics[scale=0.475]{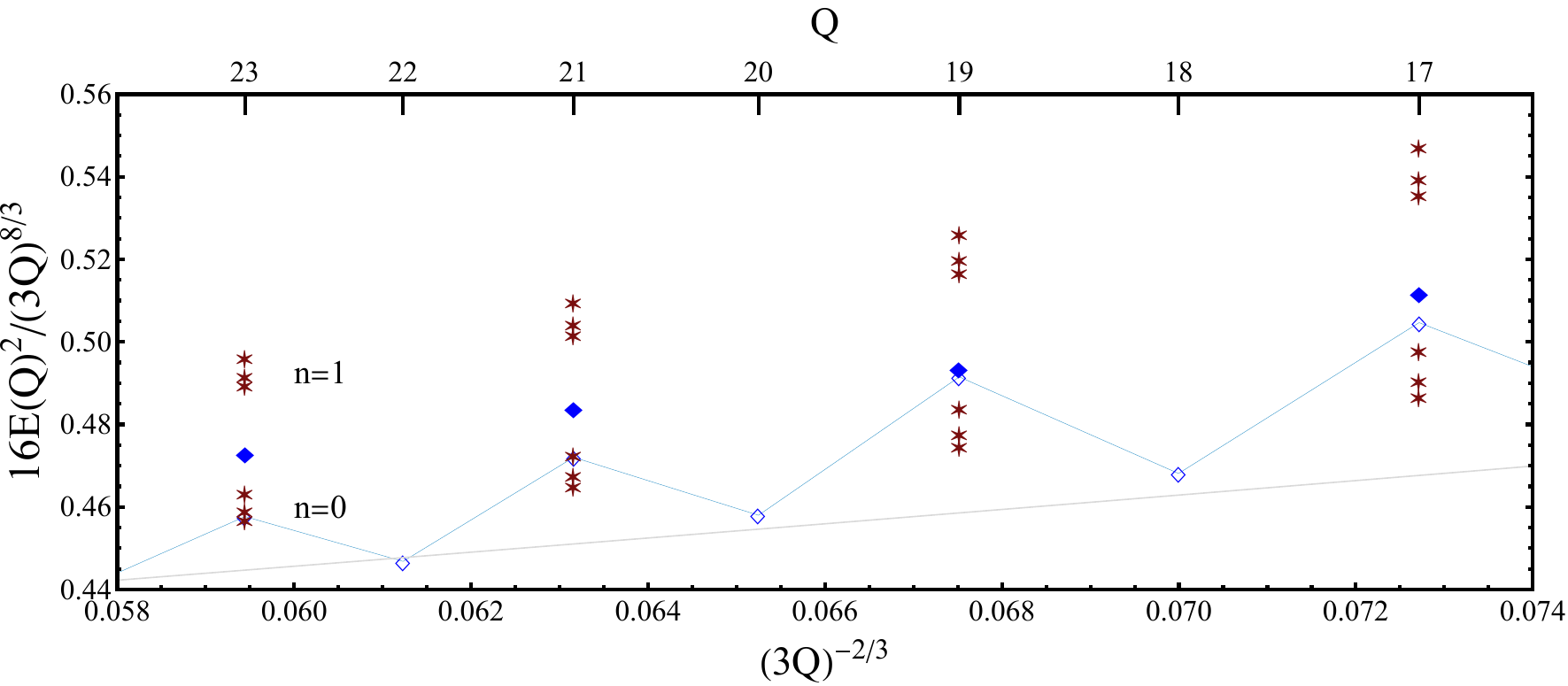}    
    \end{tabular}
    \caption{Above: prediction (dard red stars) with propagated uncertainties compared to FN-DMC-B data, including excited levels (filled blue diamonds), and to FN-DMC-2007 data.
      Below: same but with excited radial and rotational levels (uncertainties omitted for clarity).
       \label{fig:EvsQgs}}
\end{center}
\end{figure}

As the \ac{qmc} data are variational and therefore provide upper bounds, the lowest-energy results are likely to be the most accurate, and therefore in Figure~\ref{fig:EvsQgs}~(above) we compare our predictions with the \textsc{fn-dmc-b} data, and, for comparison, with the \textsc{fn-dmc-2007} data. Fitting the $Q$-even data in the range $Q=10-32$ using the large-charge energy for $Q$ even to next-to-leading order~\cite{Son:2005rv,Favrod:2018xov,Kravec:2019djc,Hellerman:2021qzz,Beane:2024kld}
\begin{eqnarray}
E(Q) \; =\;  \frac{3^{4/3}}{4} \xi^{1/2}Q^{4/3}\;-\; 3^{2/3 }\sqrt{2}\pi^2 \xi\,c_1\, Q^{2/3} \; +\; \order{Q^{5/9}} \; +\; \ldots 
\label{eq:icd}
\end{eqnarray}
yields $c_1=-0.053(2)$ and $\xi=0.355(3)$,
the  latter a value of the Bertsch parameter that is consistent with the best-known value from simulation (shown in Figure~\ref{fig:WernerALL} as the fit to the AFMC-2014 data). We then compare our value of the staggering energy to the $Q$-odd simulation data (as an addition to the best-fit curve to the even data).
The uncertainties are carried over from the input parameters given in the Tables, as described above. In Figure~\ref{fig:EvsQgs}~(below) predictions are shown (without uncertainties for clarity) for higher-orbital states and an excited radial state.
Our comparison to existing simulation data\footnote{Note that the results of Ref.~\cite{Endres:2011er} have not been included as the data has large shell effects~\cite{CRPHYS_2024__25_G1_179_0,PhysRevA.86.013626}. The results of Ref.~\cite{Mukherjee_2013} have not been included as we were unable to acquire the raw data with uncertainties. Note however that these data are at higher energy than the sets that we compared our results with.} should be interpreted primarily as a consistency check of the edge-quasiparticle picture and of the predicted large-$Q$ scaling, rather than as a precision determination of finite-$Q$ odd-particle energies.

In summary, the numerical method solves the full trapped $l=0$ \ac{bdg}
problem without expanding the potential near the edge.
The edge-scaling laws are then extracted \emph{a posteriori} from the dependence of
the eigenvalues and eigenfunctions on $\mu/\omega$.
In particular, we recover the expected behavior scaling behavior of the odd-even splitting in Eq.~\eqref{eq:Son-formula} and measure the universal coefficient \(\chi\) that we identify with
\begin{equation}
  \chi = C[\mathcal{E}_{0,0}] \, .
\end{equation}
Our best value of the universal Son parameter is therefore \(\chi=0.72(2)\),  which is consistent with the value determined by a fit to the FN-DMC-2007 data~\cite{Blume_2007}.

\begin{table}[t]
\centering
\begin{tabular}{lSSS}
\toprule
 & {\(\sigma/E_F\)} & {\(\xi \equiv \mu/E_F\)} & {\(y \equiv \sigma/\mu\)} \\
\midrule
Large-$N$ (\acs{lo}) & 0.686 & 0.591 & 1.162    \\
Large-$N$ (\acs{nlo})& 0.523 & 0.279 & 1.875   \\
\ac{qmc}            & 0.45(5) & 0.372(5)  & 1.22(5)   \\
\bottomrule 
\end{tabular}
\caption{Dimensionless gap, chemical potential, and gap-to-chemical-potential ratio for the unitary Fermi gas.
  The large-$N$ results are from Ref.~\protect\cite{Veillette:2007zz,Hellerman:2023myh}.
  The QMC results for the superfluid gap
  and energy are taken from Refs.~\cite{Carlson:2007omx} and~\cite{Carlson:2014pxa}, respectively.}
\label{tab:gap-mu-y}
\end{table}

\begin{table}[b]
  \centering
\begin{tabular}{lSSSS}
    \toprule
                & {Large-$N$ \acs{lo}} & {Large-$N$ \acs{nlo}} & {\ac{qmc}}  & {theory}     \\
    \midrule
    {$C[\mathcal{E}_{0,0}]$} & 0.71 & 0.86 & 0.72(2) & {n/a}        \\
    {$p[\mathcal{E}_{0,0}]$} & 0.33 & 0.33 & 0.33    & {$1/3$}      \\[.3em]
    {$C[\mathcal{E}_{1,0}]$} & 1.67 & 1.98 & 1.71(3) & {n/a}        \\
    {$p[\mathcal{E}_{1,0}]$} & 0.33 & 0.33 & 0.33    & {$1/3$}      \\[.3em]
    {${1}/{J_0}$}            & 1.87 & 1.48 & 1.82(5) & {$>1$}        \\
    {$C[\bar r]$}            & 0.98 & 1.00 & 0.98    & 1            \\
    {$p[\bar r]$}            & 0.50 & 0.50 & 0.50    & {$1/2$}      \\[.3em]
    {$C[w]$}                 & 0.70 & 0.65 & 0.69    & {n/a}        \\
    {$p[w]$}                 & -0.16 & -0.16 & -0.16  & {$-1/6$}     \\
    \bottomrule
\end{tabular}
  \caption{Numerical fit parameters extracted from the \ac{bdg} analysis for the energy of the lowest state \(\mathcal{E}_{0,0}\), the energy of the first excited state with \(l = 0\), \(\mathcal{E}_{1,0}\), the effective mass \(M_{\parallel} = 1/J_0\), the center of the probability distribution for the lowest eigenstate \(\bar r\) and its width \(w\).
    For each quantity we report the results of a power-law fit \(C[\Op](2 \mu / \omega)^{p[\Op]}\).
    In the last column we report the theoretical prediction from the analysis in Section~\ref{sec:probe-approximation} and~\ref{sec:eft} for which we find a remarkable agreement.
  The values of \(C[\Op]\) are low-energy parameters in the EFT.}
  \label{tab:bdg-fit-parameters}
\end{table}

\begin{figure}[t]
  \begin{center}
    \includegraphics[scale=0.63]{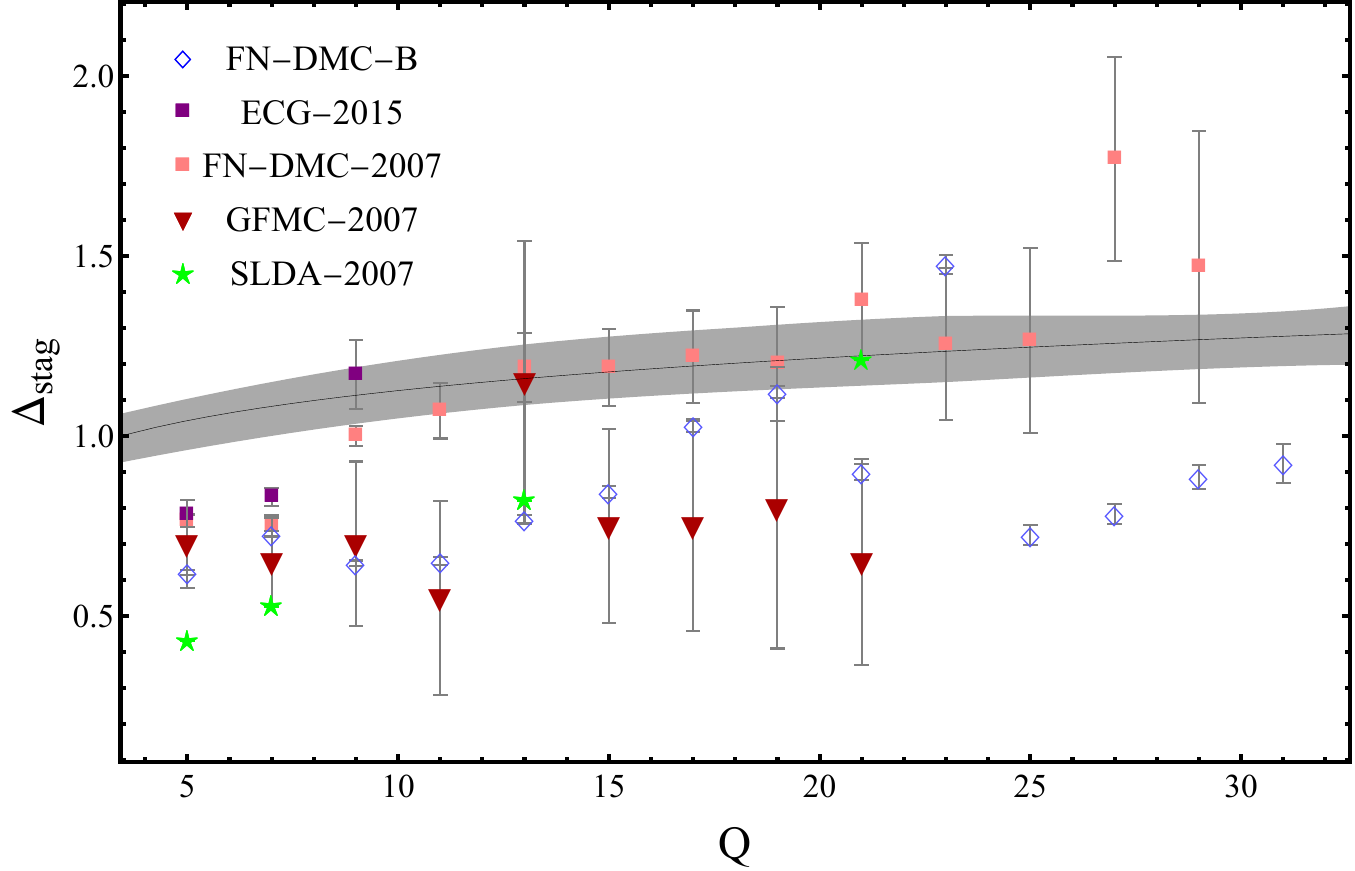}
    \caption{\small The staggering energy vs. $Q$ for the various data sets described in the text. The band is the prediction taken from the parameters in Table~\ref{tab:bdg-fit-parameters}.
       \label{fig:EstagvsQfit}}
\end{center}
\end{figure}

\FloatBarrier

\section{Conclusions}

In this note, we have studied the staggering energy at zero temperature of an unpaired fermion added to the unitary Fermi gas confined in a harmonic trap.
In~\cite{Son2007}, it was argued that at large odd particle number $Q$, the quasiparticle associated to the unpaired fermion is localized at the edge of the particle cloud and that the even-odd splitting scales as $~Q^{1/9}$.
We have followed a multi-pronged line-of-attack to verify and extend this prediction. 

We first studied the \ac{bdg} problem which is based on large-N or mean-field arguments in the probe approximation, \emph{i.e.} ignoring the backreaction of the fermion on the condensate background. Additionally, we worked in the \ac{lda} for the \ac{bdg} dynamics, and also assumed that the unpaired fermion is localized on the edge, which is justified by the observation that this is where the gap is the smallest. Like this, we were not only able to derive the scaling of the staggering gap of~\cite{Son2007}, but also derive the form of the low-lying fermion spectrum.

 Our second method was to describe our set-up with a large-charge \ac{eft}. To do so, we re-cast the edge terms introduced in~\cite{Hellerman:2020eff} into a different form and extended the usual non-relativistic superfluid description to odd particle number. Using this method, we were able to go beyond the results derived in Section~\ref{sec:probe-approximation} and in particular to capture fermion-Goldstone interactions which are frozen out by the probe approximation. The \ac{eft} method however introduces two \ac{lec}s which cannot be computed within the \ac{eft}. 
 
 Our last method was numerical. While again based on the probe approximation, it was more general in the sense that we did not need to rely on any further assumptions regarding the spectrum and the localization of the low-lying states. We found an excellent match with our results found with the other methods and gave detailed comparisons with existing \textsc{mc} data.

The comparison to many-body simulation data should be viewed primarily as a consistency check of the large-$Q$ edge picture, not as a precision determination of finite-$Q$ odd-particle energies.
The robust predictions of the present framework are:
\begin{itemize}
\item the existence of an edge-localized odd quasiparticle,
\item the scaling $\Delta_{\text{stag}}(Q)\propto Q^{1/9}$, and
\item an order-one value of the associated universal coefficient. 
\end{itemize}
By contrast, detailed finite-$Q$ energies are sensitive to the choice of bulk input parameters.

\medskip
There are a number of further directions that would be interesting to pursue. One would be to study the \ac{eft} of the unpaired fermion at the edge in more detail.

\newpage

\section*{Acknowledgments}

\begin{small}\sffamily
The authors would like to thank D.~Blume, A.~Bulgac, S.~Gandolfi, S.~Hellerman, S.~Reddy, and F.~Werner
 for useful discussions. We are particularly grateful to S.~Gandolfi for providing his unpublished data. This work was supported by the Swiss National
 Science Foundation under grant number 200021\_219267. In addition,
S.R.B is supported by the U.~S.~Department of Energy grant
DE-FG02-97ER-41014 (UW Nuclear Theory).  The authors used large language model tools to assist with code drafting, debugging, and figure preparation. All AI-assisted outputs were reviewed and verified by the authors, who take full responsibility for the final content of the manuscript.
 \end{small}

\FloatBarrier

\appendix

\section{Discrete spectrum for \(H_{\perp} \)}
\label{sec:discr-spectr-h_perp}

The numerical analysis in Section~\ref{sec:numerics} points to the fact that the orthogonal Hamiltonian \(H_{\perp}\) has a discrete spectrum.
In this appendix we want to supplement this observation with a \ac{wkb}-based semiclassical argument.

The idea is the following.
Consider the space of solutions to the eigenvalue problem
\begin{equation}
  H_{\perp} \ket{\chi(u)}  = \chi \ket{\chi(u)},
\end{equation}
where
\begin{equation}
  H_{\perp} = - \pqty*{ \frac{1}{2} \partial_u^2 + u} \sigma_3 + y u \ \Theta(u) \sigma_1 \, .
\end{equation}
Let \(\chi^+(u)\) be the solutions that decay for \(u \to \infty\) and \(S^+(\chi)\) the space of Cauchy data at \(u = 0\) that correspond to these solutions.
Similarly, let \(S^-(\chi)\) be the space of Cauchy data corresponding to solutions \(\chi^-(u)\) that decay for \(u \to - \infty\).
Normalizable eigenfunctions exist only if \(S^+(\chi)\) and \(S^-(\chi)\) have a non-zero intersection.
We want to show that both spaces have dimension two, while the total space of Cauchy data has dimension four (the eigenvalue problem is a second-order linear system for two functions).
Generically, two 2-planes do not intersect in \(\setR^4\): this can only happen for special, isolated values of \(\chi\), \emph{i.e.} the spectrum is discrete.
Note that this is a strong indication but not a full proof.
To show that the two spaces actually intersect for discrete values of \(\chi\) we would have to compute explicitly the \(S^{\pm}(\chi)\), which we cannot do analytically.

For \(u < 0\), the eigenvalue problem reduces to two decoupled Airy equations (in our approximation the gap is zero out of the droplet).
Each of the two equations admits a decaying solution that we call respectively \(\chi^-_+(u)\) and \(\chi^-_-(u)\).
Asymptotically, they are
\begin{equation}
  \chi^-_{\pm}(u) \mathrel{\underset{u \to - \infty }{\sim}} \pqty*{u \pm \chi} e^{1/3( u \pm \chi)}
  \begin{pmatrix}
    1 \pm 1 \\ 1 \mp 1
  \end{pmatrix},
\end{equation}
so \(S^-(\chi)\) has dimension two.

For \(u > 0\), the Hamiltonian is
\begin{equation}
  H^+_{\perp} = - \pqty*{ \frac{1}{2} \partial_u^2 + u} \sigma_3 + y u  \sigma_1 \, .
\end{equation}
In this case we don't have an exact solution, but we just need to study the behavior for \(u \to \infty\).
In this limit we can use a \ac{wkb} expansion and write
\begin{equation}
  \chi^+(u) = e^{S(u)} w(u)  ,
\end{equation}
where \(w(u)\) is a slowly-varying function.
Keeping only the leading term in the \(u \to \infty\) expansion, the eigenvalue equation reduces to
\begin{equation}
    \bqty*{ - \pqty*{ \frac{1}{2}\pqty*{S'}^2 + u}  \sigma_3 + y u \sigma_1} w = 0 \, .
\end{equation}
The natural ansatz is \(\pqty*{S'}^2 =2 q u\), and the problem reduces to
\begin{equation}
  \bqty*{ \pqty*{q^2 + 1} \sigma_3 - y u \sigma_1 } w = 0  ,
\end{equation}
which admits the solutions
\begin{align}
  w_{\pm} =
  \begin{pmatrix}
    1 \\ \pm i
  \end{pmatrix}, &&
  q_{\pm} = - 1 \pm i y.
\end{align}
These correspond to the two independent decaying solutions for the initial problem
\begin{equation}
  \chi^+_{\pm} \mathrel{\underset{u \to \infty }{\sim}} e^{-1/3 \sqrt{-1 \pm i y} u^{3/2}}
  \begin{pmatrix}
    1 \\ \pm i
  \end{pmatrix} \, .
\end{equation}
As advertised, also \(S^+(\chi)\) has dimension 2 and generically it will intersect \(S^-(\chi)\) only for special values of \(\chi\).

\setstretch{1}

\printbibliography

\end{document}